\documentstyle[aas2pp4, tighten]{article}

\newcommand{\xray}{\hbox{X--ray}}

\newcommand{\einstein}{Einstein}
\newcommand{\ginga}{Ginga}
\newcommand{\etal}{et al.}

\slugcomment{To appear in the Astrophysical Journal}

\begin{document}

\lefthead{Gotthelf, Hamilton, \& Helfand (1996)}
\righthead{The Detection of Faint X-Ray Flashes}

\title{\bf The EINSTEIN Observatory Detection of Faint X-Ray Flashes }

\author{E.\ V.\ Gotthelf\altaffilmark{1}}
\affil{Laboratory for High Energy Astrophysics}
\affil{NASA/Goddard Space Flight Center, Greenbelt,~MD 20771}
\affil{Electronic Mail: gotthelf@gsfc.nasa.gov}

\author{T. T. Hamilton}
\affil{Department of Astronomy}
\affil{California Institute of Technology, Pasadena,~CA 91125}
\affil{Electronic Mail: tth@astro.caltech.edu}

\and

\author{D. J. Helfand}
\affil{Department of Astronomy and Columbia Astrophysics Laboratory}
\affil{Columbia University, 538 West 120th Street, New York,~NY 10027}
\affil{Electronic Mail: djh@carmen.phys.columbia.edu}

\altaffiltext{1}{\rm Universities Space Research Association}

\bigskip\bigskip\bigskip

\centerline{To appear in the Astrophysical Journal}

\clearpage

\begin{abstract}


\doublespace

We report on the result of an extensive search for \xray\ counterparts
to Gamma Ray Bursts (GRB) using data acquired with the Imaging
Proportional Counter (IPC) on-board the \einstein\ Observatory. We
examine background sky fields from all pointed observations for
short timescale ($\lesssim 10$ sec) transient X-ray phenomena not associated
{\it a-priori} with detectable point sources. A total of $ 1.5 \times 10^7$
seconds of exposure time was searched on arc-minute spatial scales
down to a limiting sensitivity of $10^{-11}$~erg cm$^{-2}$~sec$^{-1}$
in the $0.2 - 3.5$~keV IPC band. Forty-two highly significant \xray\
flashes (Poisson probability $< 10^{-7}$ of being produced by
statistical fluctuations) are discovered, of which eighteen have
spectra consistent with an extragalactic origin and lightcurves
similar to the \ginga-detected \xray\ counterparts to GRB. Great care
is taken to identify and exclude instrumental and observational
artifacts; we develop a set of tests to cull events which may be
associated with spacecraft or near-Earth space backgrounds. The
flashes are found to be distributed isotropically on the sky and have
an approximately Euclidean number-size relation. They are not
associated with any known sources and, in particular, they do not
correlate with the nearby galaxy distribution. Whether or not these
flashes are astrophysical and/or associated with GRB, the limits
imposed by the search described herein produces important constraints
on GRB models. In this paper, we discuss possible origins for these
flashes; in a companion paper, Hamilton, Gotthelf, \& Helfand (1996) we
use the results of our search to constrain strongly all halo models
for GRB.

\end{abstract}

\singlespace

\keywords{gamma rays: bursts - surveys - X-rays: bursts}

\section {Introduction}

	Observational evidence for X-ray counterparts to classical GRB
in the 0.1 -- 100 keV range has been found by several
experiments. \xray\ bursts in the 6 -- 150 keV band are located to
one-degree precision with the WATCH all-sky monitor on-board the
GRANAT Observatory at a rate of $\sim 4$ per year per 4 pi steradians
(Lund, Brandt, \& Castro-Tirado, 1991). \ginga\ has detected X-ray bursts
associated with GRB in the 1 -- 10 keV band with similar spatial
resolution (Yosida \etal\ 1989) and has provided evidence for X-ray
precursors to GRB (Murakami \etal\ 1991). The XMON instrument on the
P78-1 mission also detected coincident \xray\ emission in this band
(Laros \etal\ 1984). As with most GRB observations, these experiments,
with their broad sky coverage and high background, offer limited
spatial resolution and detection sensitivity, leaving open the origin,
and thus the nature, of classical GRB (see Higdon \& Lingenfelter
1990 for a pre-GRO review). The existence of such X-ray counterparts
raises the possibility that they could be used to gain more detailed
positional information about the associated bursts, constrain models
for the burst number as a function of sensitivity, and, of course,
constrain directly models of the underlying physical phenomenon. To
this end, the imaging X-ray experiment on-board the \einstein\
Observatory is ideally suited to search for GRB with unprecedented
sensitivity and spatial resolution.

	The \einstein\ IPC operated for three years, producing images of
the X-ray sky with arc-minute resolution over a one square degree
field of view (FOV). Considering the operating efficiency of the
Observatory, this is equivalent to observing the whole sky for 7
minutes. It is therefore not surprising that \einstein\ was not
coincidentally pointed at any catalogued GRB which occurred during its
mission. However, by searching all images for serendipitous flashes
which occurred in portions of the field of view not occupied by any
detectable point source, we can locate all transient events with a
fluence greater than $10^{-11}$ erg cm$^{-2}$, three orders of
magnitude lower than the \xray\ counterparts of GRB at the BATSE
detection threshold. About 3\% of the \einstein\ database was searched
previously and one potentially astronomical event was found (Helfand
\& Vrtilek 1983); we also find this event in the current search,
although we exclude it from our formal sample because it does not meet
the stringent background requirements described below. Apart from this
pioneering effort, previous standard analysis of these fields used
software which was not designed to detect such flashes.

	We present the results of our search below. We have found 42
highly significant \xray\ flashes. The sky positions of these
candidates are not correlated with sources cataloged in the SIMBAD or
NED databases, nor are the events found to be coincident in time with
known GRB or solar flares detected by the Gamma Ray Spectrometer
(GRS) aboard the Solar Maximum Mission (SMM). Most importantly,
they are not preferentially found in the direction of nearby
galaxies, allowing us to challenge the viability of all halo models
for GRB (see Hamilton, Gotthelf, \& Helfand 1996). 

In \S2, we first present necessary details of the operation of the
\einstein\ IPC and the contents of its database; we then describe our
search strategy. In \S3, we report the results of the search and
examine the temporal, spectral, geographic, celestial, and detector
distributions of the flashes, none of which can be used to preclude an
astrophysical origin for the events (detailed tests for various
non-astrophysical origins are described in Appendix B). In \S4, we
evaluate the count rates recorded by the \einstein\ MPC monitor detector
during the IPC flash times. The final section discusses the origin of
the flashes that we {\it do} detect, while our companion paper uses the
results of this search to set strong constraints on the origin of GRB
which follow from the flashes we {\it do not} detect.

\section {The Data and their Analysis}

\subsection {The {IPC} Instrument}

	To describe fully this undertaking, we must first present the
workings of the IPC and what is meant by an IPC `count'.  Details of
the \einstein\ Observatory and the IPC can be found in Giacconi et
al. (1979) and Gorenstein, Harnden, \& Fabricant (1981). A complete
description of the types of events which produce detectable counts in
the IPC and an analysis of such counts are given in Wu \etal\ (1991),
Wang \etal\ (1991), and Hamilton \& Helfand (1991). We give only a
brief introduction here and consider specific relevant issues in later
sections.

	The \einstein\ Observatory was in service from late 1978 to
mid-1981 in an equatorial ($\pm 22{\arcdeg}$) low-Earth orbit. It carried
a high resolution grazing incidence X-ray telescope with a 3.4 m focal
length and a $\simeq 1{\arcdeg} \times 1{\arcdeg}$ field of view. The IPC was
one of four focal plane instruments which could be rotated into the
optical axis. Simultaneous observations were obtained with an external
co-aligned non-imaging monitor counter (\S 4).  The IPC was a
gas-filled multiwire counter containing an Ar-Xe-CO$_2$ mixture as the
detection medium. The instrument covered an energy range of $0.1 <
\rm{E} < 4.5$ keV with a resolution of $\Delta{E/E} \simeq
0.5E^{-1/2}_{\rm keV}$. The readout system used the `rise time'
method to determine event positions to an instrument-limited
resolution of from $1^\prime$ to $3^\prime$ depending on the deposited
energy. An anti-coincidence guard counter provided on-board real-time
hardware rejection of $\gtrsim 99\%$ of the particle
background. Particles and cosmic rays which deposited all their energy
within the detection volume were rejected by their (slow) time
signatures and (extreme) pulse height signals.

	A signal is initiated in the IPC with the introduction of
ionizing radiation into the counter gas. X-rays collected and focused
by the mirror enter the detection volume through a thin plastic window
and interact with the gas via the photo-electric effect. The energy of
the resulting primary electron is proportional to the incident photon
energy, thus preserving spectral information. As the electron
traverses the gas, secondary electrons are generated through
collisionally induced ionizations. The resulting electron cloud then
drifts through an electric field defined by a set of high voltage wire
planes spaced 3 mm apart. Close to the wires, the electric field
strength is high enough ($\sim 10\ {\rm kV\ cm}^{-1}$) to generate
additional collisionally ionized electrons, increasing their numbers
exponentially. This charge avalanche lasts a few $ \mu\>{\rm sec}$ and
results in a sufficient net electronic gain to allow a charge pulse,
with its distinct rise time signature, to be measured with a charge
sensitive preamplifier. Encoded into the IPC `count' for this
interaction is the time, location, and energy of the incident
radiation.

	The X-rays imaged on the detector included those of Galactic,
extragalactic and Solar origin. The diffuse Galactic X-ray background
arises from a hot bubble of gas $\sim 100$ pc in radius which surrounds the
Sun (McCammon \etal\ 1983), a ridge of emission along the Galactic
plane (Iwan \etal\ 1982; Koyama \etal\ 1986), and a putative corona of
hot gas surrounding the Galaxy. The integrated X-ray emission of known
active galactic nuclei produces about 40\% of the extragalactic \xray\
background; the origin of the balance remains unknown. The measurement
of these backgrounds by the \einstein\ IPC is discussed extensively in
Wu \etal\ (1991). In addition, the Sun is a copious emitter of \xray s
and whenever the satellite was illuminated by solar X-rays (a large
fraction of the time), a substantial number of X-rays are scattered
into the \einstein\ telescope by the residual atmosphere of the
Earth (see Wang \etal\ 1991; Fink, Schmitt, \& Harnden 1988 and
references therein.)

	The IPC is also sensitive to non X-ray events. Low-energy
electrons, $\gamma$-rays, and cosmic rays initiate avalanche events within
the detector. High energy cosmic rays produce secondary particles and
a $\gamma$-ray background from spallation in the detector walls as well as
from neutron activation of the spacecraft mass. In addition to this
`natural' background, a low-level leak of the on-board Cm/Al
fluorescence calibration source produced detectable counts in the
higher energy channels (Harnden \etal\ 1984).  Detector anomalies such
as breakdown in the counter gas and electronic malfunctions could also
result in recorded counts.

\subsection {The IPC Database}

	The IPC data base consists of 4082 pointed observations
comprising 11,230 triplets of data files: XPR, TGR, and ASP. For each
detected count, the time-ordered XPR files list an arrival time with
63  $ \mu\>{\rm sec}$ resolution, raw and gain-corrected (PI) pulse heights
digitized in 32 channels, assigned sky and detector positions with
8$^{\prime\prime}$ per pixel binning, and instrument status
information. Satellite position and orientation information is given
in the ASP files which contain a derived entry for each major frame
(40.96 sec), and include the Sun's position, Earth-Sun angle, and the
orientation of the satellite with respect to the Earth's magnetic
field. Spacecraft and detector status is given in the TGR files in
which an updated record is written each time a status flag or
environment code changes. To reject unsuitable time intervals, the IPC
data were filtered using the TGR criteria corresponding to standard
SAO processing (see Appendix B). This resulted in 184 days of
observing time out of a total of 374 days of IPC file time.

\subsection {Search Strategy}

 From the filtered data intervals, we extracted $22 \times 10^6$
counts distributed among the 11,230 files whose average observing time
was $\sim 1400$ seconds. Photons recorded in PI bins 2 -- 10,
corresponding to a nominal energy range of 0.16 -- 3.5 keV, were used
in the search. The mean count rate was $\sim 4 \times 10^{-4}$ counts
s$^{-1}$ arcmin$^{-2}$. We binned the data into space-time cells
$4^\prime\!\!.3$ by $4^\prime\!\!.3$ by 10 seconds in volume, and
searched for cells with 5 or more counts. To exclude all discrete
sources, including the bright target object typically found in the
center of the field of view, we searched only those spatial cells with
a count rate less than $6\times 10^{-4}$~counts~s$^{-1}$
~arcmin$^{-2}$. Thus, those cells flagged as meeting our criteria
contained a minimum flux enhancement of a factor of $\sim 50$ over the
mean rate for that point in space. As a practical matter, all 2179
files with less than 400 seconds of good time were excluded, since 5
counts in one spatial cell in these short files would exceed our count
rate threshold (1084 of these excluded files contained zero events
satisfying our data editing criteria). To avoid missing flashes which
straddle two cells, we actually binned the data at twice the needed
resolution and constructed each search cell from the sum of the 8
adjacent (three-dimensional) sub-cells (i.e., we searched all possible
overlapping cells).

	A total of 69 events flagged in this way were then inspected
for their spatial extent and compared to a simultaneous background
count rate defined by a concentric annulus $7^{\prime}$ to
$15^{\prime}$ in radius and $\pm 20$~s in time of the detection cell
(50 s duration). 

We rejected all events that represented deviations of $< 3\>\sigma$
above this local background (i.e., in which $3 > | N_{\rm obs} -
N_{\rm exp} | / \sqrt{N_{\rm exp}}$, where $N_{\rm obs}$ is the number
of counts recorded in the spatial pixel of interest and $N_{\rm exp}$
is the predicted number from the background annulus).  Subsequent
scrutiny revealed that the events excluded by this test were associated
with satellite sunrise, sunset, or the approach of the South Atlantic
Anomaly. Furthermore, in order to exclude events possibly associated
with glitches at the beginning or end of an observation interval, we
eliminate those occurring within 20 s of gaps in the data. These cuts
left a total of 42 accepted events. Formally, we would expect less than one
event of such intensity as a result of Poisson noise.

\subsection {The Detection of X-ray Flashes} 

	The 42 X-ray flash candidates were analyzed for their spectral
and temporal structure. The data constituting an 'event' were further
characterized in the following manner. The sky position is given by
the mean position of all counts falling within $3^{\prime}$ in space and
$\pm 20$ s in time of the three-dimensional detection cell. The event
time is taken as the (first) 1-second bin containing the maximum
number of counts as determined using a 0.1 s sliding box during the 50
s time window. Each event was then classified based on its time
structure as slow or rapid, and based on its spectrum as soft or
hard. The rapid events are defined as having $> 4$ counts occurring
within the 1 s maximum bin defined above.  Events for which ${\rm
N}(< 1.3\ {\rm keV}) < {\rm N}(\geq 1.3\ {\rm keV})$ are defined as
hard. The sky position, event times, and characteristics of the events
passing the initial inspection are presented in Table 1.  Of the
42 events, 36 are slow, 6 are rapid, 18 are hard, and 24 are soft. It
is found that all but one of the hard events are slow, and that
similarly, all but one of the rapid events are soft. The converse in
both cases is not true, however: soft events are both slow (79\%) and
rapid (21\%), and slow events are composed of nearly equal parts hard
and soft events.

	Light curves are shown in Fig. 1 for four representative
flashes. The fact that the rise times are shorter than the decay times
is not an artifact of our search criteria. In Fig. 2, a composite
light curve for each temporal class is shown. Each flash has been
centered on the 1 sec bin with the most counts as described
above. Most of the accepted events have a `slow' time structure, with
90\% of the counts clustered within a window 15 s in duration compared
to 90\% window boundaries of 1 s for the rapid events. An aggregate
point spread function (PSF) using all accepted events was produced by
centering and stacking events on the computed mean sky positions. The
result is plotted in Fig. 3 along with the summed spectrum. The
resulting PSF resembles a point source and, the summed spectrum,
although quite soft, appears astrophysical (see below).  The events
were then grouped by spectral hardness and analyzed in a similar
manner.

Fig. 4 displays the PSF and spectra for the soft and hard events
separately. The soft events contain on average 11 counts per event and
produce a PSF with a normalized radial distribution consistent with a
Gaussian of ${\rm FWHM} \simeq 3^\prime$. This is the signature
expected for a point source with the observed soft spectrum imaged by
the mirror plus the IPC. The hard events average 7 counts and also
produce a radial distribution consistent with an astronomical origin.

In Fig. 5, the soft events are compared directly with data from the
cataclysmic variable U Gem in outburst, put through the same analysis;
it is clear that both the PSF and spectrum are consistent with those
of a real astrophysical source. Fig. 6 displays the properties of slow
and rapid events separately. The characteristics for the slow events
are similar to those of the sample as a whole. The six rapid events
contain $\sim$ 16 counts per event for a total of 100 counts and produce a
very soft spectrum with 75\% of the counts below $\sim 0.4$
keV. Inspection of the counts contained within those 1 sec bins which
contained $> 4$ counts reveal an even softer spectrum. Recent studies
of super-soft objects show that such spectra, although rare, are found
in nature (Brown \etal\ 1994; Greiner, Hasinger, \& Kahabka 1991;
Cowley \etal\ 1993).

A sky exposure map was produced for the complete IPC database by
summing up the filtered time intervals for each sky pointing. The map
is plotted in Galactic coordinates in Fig. 7 with the candidate flash
locations overlaid. The accepted events are roughly isotropic on the
celestial sphere. In particular, as we show in the accompanying paper
(Hamilton, Gotthelf, \& Helfand 1996), they are not found
preferentially in the direction of nearby galaxies.

A $\log N-\log S$ curve for all accepted events was constructed using
a mean conversion factor of $2.6 \times 10^{-11} \ {\rm erg \>
cm}^{-2}$ per count. We compensated for the off-axis mirror reflection
efficiency using the standard IPC vignetting correction (Harris \&
Irwin 1984). The result is plotted in Fig. 8 along with various
(pre-GRO)) measurements (Higdon \& Lingenfelter 1990). Although it is
quite possible that both the $\gamma$-ray and X-ray event curves
suffer from instrumental bias (see Higdon \& Lingenfelter 1984; Mazets
\& Golenetskii 1987 for a review of this issue as related to the {\sl
Konus} data, and Meegan \etal\ 1994 for a discussion of the BATSE
results), a naive straight line between the two curves gives a
differential slope of $\sim -1.7$, similar to the slope detected by
BATSE for the number-size relationship of the faintest bursts it
detects and considerably flatter than the $\sim -2.5$ slope for a
homogeneous Euclidean distribution.

	As discussed above, the flashes exhibit a variety of temporal
and spectral characteristics. These effects are not uncorrelated, the
softest flashes often having the fastest rise times. Although all 42
flashes taken together do not show any correlation with position in
the detector, when we examine only the soft-rapid flashes we find that
their positions in physical detector coordinates are preferentially
aligned along the directions of the counter wires. While it is by no
means excluded that some regions of the counter may have heightened
sensitivity to genuine astronomical events, the observed pattern
suggests that these events may originate within the detector. One
possible explanation is that these events are afterpulses, counter
events which result when not all of the electrons from a rejected
particle event fall within the effective counter dead time; i.e., one
or a few slow moving electrons may not have reached the counter wires
until after the counter has reset following the event which generated
them. These events have been well-studied in the ROSAT PSPC (Snowden
\etal\ 1994), and it is very unlikely that groups of afterpulses would
appear as discrete counts as do the events which make up the flashes
(Snowden, private communication). However, we consider it likely that
the softer and/or more rapid flashes detected by our search are related
to afterpulses or some other counter phenomenon and we excluded them
from further consideration. The fact that we thereby delete 24 events
as possible counter artifacts does not imply that we necessarily
believe the remaining 18 events with slower rise times and harder
spectra to be true astronomical events.

\section {MPC Results}

The \einstein\ Monitor Proportional Counter (MPC) was an independent
non-focal plane instrument co-aligned with the optical axis of the
\einstein\ observatory. The MPC was a collimated proportional counter
with a 1.5 mil Beryllium window which was sensitive to 1 to 20 keV
X-rays. During normal operation, the MPC simultaneously observed the
entire field of view being observed by the IPC. Thus while the flashes
we observe with the IPC may, if they have a hard spectrum, be
observable with the MPC, the thousand-fold increase in effective
background in the MPC would make detection of transients at these flux
levels impossible. However, by summing the MPC data for the 18 hard
flashes we detect we may test whether the events are astronomical and
also constrain their spectrum.

MPC data readily available from the High Energy Astrophysics Science
Archive (HEASARC) at Goddard Space Flight Center report the total MPC
counts in 2.6 second bins. A typical bin in a field with no bright
sources, such as those examined in our IPC investigation, contains
about 40 counts. We compared the counts in the time bin in which we
detect flashes to an expected background equal to the mean of the four
preceding and four following bins. Of the l8 hard events, which one
might reasonably expect to have MPC counterparts, 3 had no MPC data,
one came at the first bin of an MPC observation and one corresponded
to a mean MPC rate of over 100 counts per second. Excluding those five
events leaves 13 hard events with good MPC data. The mean MPC rate was
45, the extremes 64 and 37. None of the MPC intervals corresponded to
a count rate excursion of more than 14, or about 2 sigma. In four of
the intervals the count rate at the time of the flash was less than or
equal to the expected background. In ten of the intervals the count
rate was higher. The total counts in the thirteen 2.6 second bins
corresponding to the flash peaks was 647. This is 2.5 sigma above the
expected background of 587. This is encouraging but hardly a
definitive confirmation of an astronomical origin for the flashes.

	The mean fluence per MPC flash is therefore 4.6 photons,
although this number is probably an underestimate of the true fluence
as a result of the restrictive time bin used here. If the MPC events
had the same time signature as the IPC flashes our procedure would
detect about half of the total fluence.  The hard IPC flashes
themselves have a mean fluence of 7 photons. The ratio between MPC and
IPC is consistent with a 1.7 keV thermal plasma or a power law with a
photon index of 2. This is reasonably near the observed \ginga\ GRB
X-ray spectrum.

\subsection {Discussion} 

	Our search for X-ray transients found a total of 42 in the
\einstein\ database. This result is a robust upper limit on transient
X-ray phenomena at the detected flux levels.  However, the origin of
the signals which we do detect remains a mystery. While we have been
unable to provide a definitive explanation, several possibilities are
evident. The least interesting is that the flashes are a result of
some hitherto unknown phenomenon which takes place entirely within the
detector. As discussed above, it may be that many (or even all) of the
flashes are the result of a phenomenon analogous to the afterpulses
detected in ROSAT (Snowden \etal\ 1994). Of course it is unlikely that
such a phenomenon would manifest itself with a point response function
and spectrum which perversely imitate incident cosmic X-rays, but this
possibility cannot be excluded. A second obvious explanation is that
the flashes originate outside the spacecraft but within the solar
system. The fact that the flashes do not correlate with solar-related
phenomena such as the illumination of the spacecraft by the sun
discourages, but does not refute, such an explanation. We have
performed a number of tests to determine if the flashes correlate with
any non-astronomical phenomenon. The tests are described in
Appendix A. We conclude from these tests that an astronomical origin
for the flashes cannot be ruled out. Indeed, some of the properties of
the flashes support particular types of astronomical interpretations.

	The isotropic distribution of flashes on the celestial sphere
indicates that they originate at distances either large or small
compared to the length scale of the Galaxy ($\sim 10$ kpc). The homogeneous
differential slope of $-2.5$ exhibited by the flash $\log N-\log S$
relation is consistent with either an extragalactic or a local origin
for the flashes. While the time structure, space distribution and
apparent luminosity function of the X-ray flashes is reminiscent of
GRB, the fluence of the observed flashes is several orders of
magnitude fainter. If the flashes originate at a distance of 1 pc, the
required energy is $\sim 10^{28}$ ergs, equivalent to the energy released by
the accretion of $\sim 10^{9}$ grams of material onto a compact object. If the
flashes originate at 1 Gpc, then a typical energy would be $\sim 10^{46}$ ergs.

	Although our experiment is sensitive to flashes on a timescale
of up to $\sim 10$ seconds, the flashes we detect characteristically have a
substantially shorter timescale; indeed the flashes have a timescale
shorter than the X-ray counterparts of GRBs detected by \ginga,
especially if one postulates an origin at cosmological distances which
would require substantial time dilation in the faintest (most distant)
bursts. They are not obviously shorter than the X-ray timescale of the
SGR detected by ASCA, although the X-ray light curve of the ASCA SGR
is not well constrained because the ASCA gas scintillators saturated
their telemetry buffer after only 5 events (Murakami \etal\ 1994). We
note that the flashes we detect cannot be counterparts to cosmological
GRB if the latter are standard candles however, since, for a faint
BATSE burst redshift of $\sim 1 - 2$, there is insufficient volume of
Universe at greater distances to produce events 30 times fainter with
the frequency we observe.

	The flashes are also far more numerous than known GRB: we
observe a rate of $2 \times 10^{6}\ {\rm yr}^{-1}$ over the whole sky. This is
approximately the frequency with which supernovae occur within 1 Gpc
which prompts the speculation that the X-ray flashes are produced by
the breakout of supernova shocks as they reach the optically thin
regions of exploding stars. This would require the release of $\sim 10^{-7}$
of the supernova's energy in a prompt ultraviolet/soft X-ray flash. It
is clear, however, that not all of the observed flashes originate
beyond the Galaxy, since one soft event occurs at a Galactic latitude
of $2\arcdeg$.

	The isotropic nature of the flashes' distribution is of course
consistent with an origin relatively nearby in the galaxy. It is
tempting to propose as progenitors local flare stars and/or RS CVn
systems. The soft spectra of the detected flashes indeed suggest a
coronally active star population. Given the flare frequency and
duration of a typical flare star, there is a reasonable chance to
catch a flare event during a typical IPC observation.  But again, the
number of flashes are inconsistent with local field star
densities. Perhaps the flashes are associated with a Galactic
population of old, isolated neutron stars which is otherwise
invisible.

	Of course there is no reason to believe that the 42 events we
have detected, or even the 18 with harder spectra and slow timescales,
have a common origin. They may represent a maddening combination of
interesting astronomical events and indistinguishable counter
artifacts. In any case, the observed flash rate is a definite upper
limit on the occurrence of faint astronomical X-ray transients. The
significance of this limit as related to GRB is explored in Hamilton,
Gotthelf, \& Helfand (1996).

\acknowledgements

	DJH acknowledges support from NASA grant NAS5-32063. TTH
acknowledges support from NASA grant NAGW-4110. Part of this research
has made use of data obtained through the High Energy Astrophysics
Science Archive Research Center Online Service, provided by the
NASA-Goddard Space Flight Center. This paper is contribution no. 544
of the Columbia Astrophysics Laboratory.

\clearpage

\onecolumn


\appendix

\centerline{APPENDIX A}

\section{Tests for a non-astrophysical origin of the IPC flashes}

	In order to determine if the detected flashes truly have an
astrophysical origin, we devised a series of tests based on coverage
and sensitivity arguments.

	One possible non-astrophysical origin for the observed flashes
is arcing associated with a particular spot on a counter wire. If this were
the cause of our events, the point response function should differ
significantly from that observed for real celestial sources. Our
search box is much larger than the IPC point response function (18
vs. $\lesssim 7$ arcmin$^{2}$) so a statistical noise fluctuation
would also look very different. In addition, the spectrum from
instrumental artifacts would not show the familiar effects of counter
window absorption (mainly carbon at 0.28 keV). As mentioned in \S3,
the PSF resembles a point source and the spectrum appears
astrophysical.

	The gain and other counter properties varied substantially
during the life of the satellite so we knew that if occurrence of
transients was concentrated in one part of the mission, the cause was
probably instrumental. Thus we plotted the time of occurrence for
events during the mission. The result is shown in Fig. 9 along with
the daily fraction of filtered IPC time coverage (total time intervals
passing the standard processing criteria per day). A
Kolmogorov-Smirnov (K -- S) test yields a probability of 78\% that
the distribution is random with respect to spacecraft lifetime
(Fig. 10). We also calculated the distribution of events with respect
to local solar time. The total filtered IPC observing time was binned
into hourly intervals and compared to the event distribution. An
excess of occurrences on the leading side of the Earth's orbital
motion would suggest X-rays from meteoritic material interacting with
the Earth's atmosphere. No such effect is seen (Fig. 11).

	Nuclear reactors aboard Soviet spy satellites are a potential
source of non-astrophysical $\gamma$-ray and X-ray transients. Recently
declassified results from the SMM GRS experiment have detailed the
cause and properties of these events (Rieger \etal\  1989; Share et
al. 1989; Hones \& Higbie 1989; O'Neill \etal\ 1989). SMM and
\einstein\ flew concurrently and in similar orbits for about a year from
14 February 1980 onwards. During the overlap of the two missions, an
average of 3 anomalous events per month were detected by SMM. We
compared the monthly event rates given by Rieger \etal\ (1989 -- see
Fig. A4) to that of our flashes (normalizing by coverage) and do not
see a significant correlation. However, this is not a strong test
because the data set is too sparse, and the rate is dependent on
altitude and orbit. We looked for a periodic recurrence rate relative to
the \einstein\ orbit by plotting the time of day of the flash
occurrence against day number (see Rieger \etal\ 1989). Again no
obvious trend is evident.

	The ASP file data, filtered with the TGR data, were used to
construct a satellite Earth coverage map. The geographic location of
the satellite at the time of the events is shown in Fig. 12. The
shading represents the time that the satellite spent operating at the
designated latitude and longitude, e.g., the unshaded area in the
southern hemisphere corresponds to the South Atlantic Anomaly. No
correlation with latitude or with putative nuclear test sites is
found.  

A detector exposure map was made for our flash search by summing the
filtered times passing both the maximum count rate criteria and the
minimum count criteria as a function of detector coordinates
(Fig. 13). As expected, the exposure in the center of the counter is
suppressed because of the presence of target sources in the field. The
outline of the window support ribs is also apparent and is due to two
competing effects. Because of the geometric blockage of the various
X-ray background contributions, the low count rate exposure is
enhanced under the ribs, but there, the chance of falling below the
minimum count threshold is increased, and thus the exposure is
decreased overall.  The region with the most exposure lies just
outside the ribs where vignetting decreases the background count rate
and increases the sensitivity of our search.

We plotted the flashes in detector coordinates and found a generally
smooth distribution with an enhancement towards the edge of the field
(Fig. 14). This is consistent with our exposure map. In Fig. 14 we
outline the location of the window support ribs and the nominal window
edge. The size of the ribs is energy dependent, so that a flash can
appear to be `under' the rib, particularly since the PSF is relatively
large at low energies. We used a definition for the rib width of
$4^\prime\!.3$. Our detector exposure map includes all the available
area, including area outside that used in the standard SAO processing,
where the quality of the detector linearization, thus the aspecting,
is rather poor. Thus two flashes appear to be just outside the detector
window. However, again, this is consistent with the exposure.

To test this qualitative impression of consistency, we created a 
histogram of exposure time intervals and calculated the
number of flashes expected in each based on the total number of
flashes observed. The expected and actual distributions based on the
exposure time accumulated at various points on the detector were
consistent at the 66\% confidence level ($\chi^{2}$ value of 10.4 for
13 degrees of freedom).

Our somewhat disappointing conclusion after
this evaluation is that the possibility that the transient events are
of astrophysical origin cannot be established or excluded. If they
are celestial events, their isotropy is consistent with an origin at
the sites of the GRB; however other possibilities, such as origins at
sites in the Oort Cloud or a local Galactic population cannot be
excluded.  

An expanded version of this discussion may be found in Gotthelf (1992).


\bigskip 

\centerline{APPENDIX B}

\section{IPC data selection criteria}

The Columbia IPC database (Helfand \etal\ 1996) allows the user to
tune data selection based on the background levels, energy ranges,
aspect quality, and other instrumental and environmental factors most
appropriate for the task at hand.  The eleven TGR criteria used to
accept or reject data are displayed in Table B1 along with the range
of values allowed and the specific criteria for the transient search
reported here.


\bigskip\bigskip

\hfil\vbox{
\halign{#\hfil \tabskip 3.5em & \hfil#\hfil &
\hfil#\hfil\tabskip=0pt\cr
\multispan3\hfil{ Table B1}\hfil
\cr
\multispan3\hfil{Definition of the Standard IPC TGR Criteria}\hfil
\cr
\noalign{\vskip 1em\hrule\vskip 2pt\hrule\vskip 1em}
TGR & Value & Search\cr			
Criteria & Range & Settings\cr
\noalign{\vskip 1em\hrule\vskip 2em}					
background level & 0--4 & 0--2\cr
viewing geometry & 1--5 & 1--3\cr		
high voltage value & 0--9 & 4--9\cr
aspect separation & 0--15 & 0--15\cr
aspect solution & &\cr
{\quad no aspect mode} & on/off/both & off\cr
{\quad lock on mode} & on/off/both & on\cr
{\quad extrapolated mode} & on/off/both & off\cr
{\quad mapping mode} & on/off/both & on\cr
telemetry quality & good/bad/both & good\cr
data quality & good/bad/both & good\cr
high voltage & on/off/both & on\cr
filter status & in/out/both & out\cr
calibration mode & on/off/both & off\cr
PI bin & 0--15 & 2--10\cr
\noalign{\vskip 1em\hrule\vskip 2em}
}}\hfil

\vfill


\twocolumn


\onecolumn
\clearpage


\begin{deluxetable}{r r r c c c}

\tablenum{1}

\tablecaption{A list of faint IPC flashes
\label{tab:var-short}}

\tablehead{
\colhead{Event Time} & \colhead{R.A.} & \colhead{Declination} &
\colhead{Spectral\tablenotemark{a}} &
\colhead{Temporal\tablenotemark{a}} &
\colhead{Offset\tablenotemark{b}} \\
\colhead{(UT)} & \colhead{(B1950)} & \colhead{(B1950)} &
\colhead{Hardness} & \colhead{Structure} & \colhead{Angle} 
}

\startdata
 01/12/79   05:32:45 & 00$^h$ 44$^m$ 51$^s$ & $ 41^{\circ}$ 46$^{\prime}$ 20$^{\prime\prime}$ & soft & slow &  27$'$ \nl
 01/22/79   10:57:34 & 14$^h$ 59$^m$ 38$^s$ & $ 21^{\circ}$ 25$^{\prime}$ 29$^{\prime\prime}$ & soft & slow &  33$'$ \nl
 01/24/79   00:40:21 & 13$^h$ 52$^m$ 43$^s$ & $ 05^{\circ}$ 35$^{\prime}$ 03$^{\prime\prime}$ & soft & slow &  31$'$ \nl
 02/26/79   12:05:12 & 04$^h$ 05$^m$ 45$^s$ & $-72^{\circ}$ 04$^{\prime}$ 13$^{\prime\prime}$ & hard & slow &  42$'$ \nl
 04/10/79   17:23:16 & 05$^h$ 25$^m$ 35$^s$ & $-70^{\circ}$ 47$^{\prime}$ 55$^{\prime\prime}$ & soft & slow &  33$'$ \nl
 05/25/79   21:09:21 & 23$^h$ 02$^m$ 54$^s$ & $-23^{\circ}$ 21$^{\prime}$ 53$^{\prime\prime}$ & hard & slow &  44$'$ \nl
 08/03/79   23:25:48 & 03$^h$ 17$^m$ 25$^s$ & $ 42^{\circ}$ 12$^{\prime}$ 18$^{\prime\prime}$ & hard & slow &  42$'$ \nl
 08/13/79   20:19:00 & 15$^h$ 42$^m$ 20$^s$ & $ 21^{\circ}$ 01$^{\prime}$ 42$^{\prime\prime}$ & soft & rapid &  39$'$ \nl
 08/29/79   05:51:29 & 23$^h$ 58$^m$ 22$^s$ & $ 79^{\circ}$ 11$^{\prime}$ 19$^{\prime\prime}$ & hard & slow &  45$'$ \nl
 08/29/79   15:40:52 & 16$^h$ 00$^m$ 41$^s$ & $ 15^{\circ}$ 28$^{\prime}$ 08$^{\prime\prime}$ & hard & slow &  36$'$ \nl
 10/04/79   23:41:55 & 05$^h$ 40$^m$ 53$^s$ & $ 49^{\circ}$ 28$^{\prime}$ 13$^{\prime\prime}$ & soft & rapid &  30$'$ \nl
 10/06/79   21:22:56 & 19$^h$ 32$^m$ 51$^s$ & $ 21^{\circ}$ 28$^{\prime}$ 49$^{\prime\prime}$ & soft & rapid &  39$'$ \nl
 10/19/79   23:56:37 & 06$^h$ 00$^m$ 41$^s$ & $-40^{\circ}$ 09$^{\prime}$ 24$^{\prime\prime}$ & hard & slow &  18$'$ \nl
 10/26/79   14:25:16 & 09$^h$ 06$^m$ 15$^s$ & $ 17^{\circ}$ 00$^{\prime}$ 04$^{\prime\prime}$ & soft & rapid &  36$'$ \nl
 11/07/79   19:55:45 & 06$^h$ 27$^m$ 11$^s$ & $-55^{\circ}$ 18$^{\prime}$ 04$^{\prime\prime}$ & hard & slow &  17$'$ \nl
 11/22/79   10:32:38 & 20$^h$ 47$^m$ 40$^s$ & $ 29^{\circ}$ 10$^{\prime}$ 41$^{\prime\prime}$ & hard & slow &  26$'$ \nl
 11/25/79   17:11:26 & 20$^h$ 10$^m$ 32$^s$ & $ 36^{\circ}$ 45$^{\prime}$ 58$^{\prime\prime}$ & soft & slow &  44$'$ \nl
 12/03/79   04:04:49 & 10$^h$ 46$^m$ 21$^s$ & $ 09^{\circ}$ 10$^{\prime}$ 06$^{\prime\prime}$ & hard & slow &  38$'$ \nl
 12/05/79   17:40:11 & 12$^h$ 11$^m$ 07$^s$ & $ 14^{\circ}$ 25$^{\prime}$ 33$^{\prime\prime}$ & hard & slow &  32$'$ \nl
 12/15/79   10:16:33 & 23$^h$ 46$^m$ 40$^s$ & $-28^{\circ}$ 22$^{\prime}$ 23$^{\prime\prime}$ & hard & rapid &  39$'$ \nl
 12/16/79   15:25:51 & 12$^h$ 16$^m$ 42$^s$ & $ 30^{\circ}$ 12$^{\prime}$ 37$^{\prime\prime}$ & soft & slow &  16$'$ \nl
 01/17/80   05:59:19 & 02$^h$ 36$^m$ 16$^s$ & $-08^{\circ}$ 43$^{\prime}$ 57$^{\prime\prime}$ & soft & slow &  37$'$ \nl
 01/17/80   05:57:09 & 02$^h$ 35$^m$ 42$^s$ & $-08^{\circ}$ 24$^{\prime}$ 44$^{\prime\prime}$ & soft & slow &  43$'$ \nl
 02/05/80   13:28:17 & 12$^h$ 05$^m$ 23$^s$ & $-52^{\circ}$ 42$^{\prime}$ 58$^{\prime\prime}$ & soft & slow &  34$'$ \nl
 02/12/80   15:15:56 & 15$^h$ 50$^m$ 29$^s$ & $-23^{\circ}$ 22$^{\prime}$ 25$^{\prime\prime}$ & soft & slow &  31$'$ \nl
 03/25/80   20:00:19 & 06$^h$ 40$^m$ 06$^s$ & $ 09^{\circ}$ 59$^{\prime}$ 26$^{\prime\prime}$ & soft & slow &  32$'$ \nl
 04/09/80   14:17:18 & 07$^h$ 37$^m$ 36$^s$ & $ 18^{\circ}$ 01$^{\prime}$ 48$^{\prime\prime}$ & soft & slow &  36$'$ \nl
 04/13/80   20:36:16 & 05$^h$ 00$^m$ 39$^s$ & $-69^{\circ}$ 35$^{\prime}$ 37$^{\prime\prime}$ & soft & slow &  39$'$ \nl
 04/14/80   01:37:18 & 19$^h$ 20$^m$ 41$^s$ & $-00^{\circ}$ 13$^{\prime}$ 45$^{\prime\prime}$ & hard & slow &  32$'$ \nl
 04/14/80   10:34:24 & 17$^h$ 02$^m$ 13$^s$ & $ 60^{\circ}$ 56$^{\prime}$ 24$^{\prime\prime}$ & soft & slow &  15$'$ \nl
 04/16/80   11:26:08 & 00$^h$ 50$^m$ 10$^s$ & $-72^{\circ}$ 59$^{\prime}$ 20$^{\prime\prime}$ & soft & slow &  32$'$ \nl
 07/08/80   03:16:11 & 12$^h$ 33$^m$ 13$^s$ & $ 15^{\circ}$ 50$^{\prime}$ 19$^{\prime\prime}$ & soft & slow &  33$'$ \nl
 07/08/80   05:18:58 & 14$^h$ 13$^m$ 36$^s$ & $ 00^{\circ}$ 54$^{\prime}$ 38$^{\prime\prime}$ & hard & slow &  37$'$ \nl
 07/12/80   14:27:10 & 13$^h$ 07$^m$ 06$^s$ & $-00^{\circ}$ 53$^{\prime}$ 22$^{\prime\prime}$ & hard & slow &  28$'$ \nl
 08/07/80   22:18:57 & 15$^h$ 27$^m$ 02$^s$ & $ 12^{\circ}$ 09$^{\prime}$ 49$^{\prime\prime}$ & hard & slow &  33$'$ \nl
 08/20/80   20:56:42 & 16$^h$ 03$^m$ 19$^s$ & $ 18^{\circ}$ 42$^{\prime}$ 30$^{\prime\prime}$ & soft & slow &  22$'$ \nl
 08/20/80   21:56:08 & 16$^h$ 01$^m$ 05$^s$ & $ 18^{\circ}$ 13$^{\prime}$ 11$^{\prime\prime}$ & soft & rapid &  37$'$ \nl
 03/01/81   07:29:46 & 18$^h$ 41$^m$ 48$^s$ & $ 20^{\circ}$ 06$^{\prime}$ 05$^{\prime\prime}$ & soft & slow &  34$'$ \nl
 03/18/81   06:23:56 & 04$^h$ 46$^m$ 11$^s$ & $ 11^{\circ}$ 10$^{\prime}$ 15$^{\prime\prime}$ & hard & slow &   7$'$ \nl
 04/01/81   18:21:48 & 17$^h$ 32$^m$ 15$^s$ & $-12^{\circ}$ 35$^{\prime}$ 03$^{\prime\prime}$ & soft & slow &  41$'$ \nl
 04/03/81   16:54:50 & 17$^h$ 33$^m$ 46$^s$ & $-08^{\circ}$ 13$^{\prime}$ 33$^{\prime\prime}$ & hard & slow &  31$'$ \nl
 04/25/81   14:56:47 & 09$^h$ 27$^m$ 44$^s$ & $ 06^{\circ}$ 00$^{\prime}$ 05$^{\prime\prime}$ & hard & slow &  30$'$ \nl 
\enddata

\tablenotetext{a}{See \S 3 for definitions.}
\tablenotetext{b}{Distance of the flash from the detector center.}

\end{deluxetable}

\clearpage

\begin{figure} \figurenum{1} \caption{Light curves for four typical flash
events.  The flashes in $1a$ and $1b$ are categorized as slow and
those in $1c$ and $1d$ as rapid. The central strip in each plot shows
the arrival times of the photons as dots. The bottom strip shows the
times of housekeeping status changes as vertical lines. This indicates
that the flashes are not associated with detectable changes of
spacecraft or instrument state.
\label{Fig. 1}}
\end{figure}

\begin{figure} 
\figurenum{2} 
\caption{The composite light curves for the 
X-ray flashes are shown for both slow (right panel) and rapid (left
panel) events.  Both curves show a rise time significantly shorter
than the decay time. Since our search method is symmetric with respect
to time, this asymmetry is a real effect.
\label{Fig. 2}}
\end{figure}

\begin{figure} \figurenum{3} \caption{A composite radial distribution 
and spectrum for the 42 flashes. One pixel is eight arc-seconds. The
radial distribution is consistent with the point response function of
the IPC.
\label{fig3}}
\end{figure}

\begin{figure} \figurenum{4} \caption{A composite radial distribution and 
spectrum for each spectral classification. Both radial distributions
are consistent with a point source origin.
\label{fig4}}
\end{figure}

\begin{figure} \figurenum{5} \caption{A comparison of the composite radial 
distribution and spectrum of the soft flashes and that of U Geminorum
in outburst analyzed in the same manner. The U Geminorum outburst was
one of the softest celestial source detected by the IPC. This plot
contains 200 secs of data from the $\rm{HUT}\#1474851 + 4100$ secs.
The similarity of the spectra demonstrates that, although the soft
flashes do have an extremely soft spectrum, it is a spectrum
consistent with a celestial origin of the photons.
\label{fig5}}
\end{figure}

\begin{figure} \figurenum{6} \caption{Composite radial distributions and 
spectra sorted by temporal class. In the bottom two panels, only
events occurring in the brightest one second of the rapid flashes are
included.
\label{fig6}}
\end{figure}

\begin{figure} \figurenum{7} \caption{The location on the celestial sphere 
of the 42 X-ray flashes overlaid on an IPC sky exposure map. The
symbol size for each exposure is proportional to the duration of the
exposure. The flashes are clustered in areas of the sky, such as the
LMC, which were observed for long periods. The density of flashes per
unit exposure time does not correlate with Galactic latitude or the
locations of any known class of object.
\label{fig7}}
\end{figure}

\begin{figure} \figurenum{8} \caption{The Log $N$ -- Log $S$ relationship 
for the X-ray flash fluxes reported in this paper plotted along with
the Log $N$ -- Log $S$ relations found by various others experiments
(Higdon \& Lingenfelter 1990). 
\label{fig8}}
\end{figure}

\begin{figure} \figurenum{9} \caption{Daily IPC time coverage including 
only the time intervals passing the standard processing criteria (see
Table B1). The times of occurrence of the 42 flashes are indicated
along the abscissa.
\label{fig9}}
\end{figure}

\begin{figure} \figurenum{10} \caption{A graphical representation of the
Kolmogorov-Smirnoff test for flash occurrence time. We see no evidence
that the flashes do not occur randomly with respect to the time in the
mission.
\label{fig10}}
\end{figure}

\begin{figure} \figurenum{11} \caption{A histogram of the number of 
flashes binned by local solar time shows no statistically significant
evidence for dependence of flash occurrence on local time. If the
flashes were associated with meteoritic infall, they should occur more
frequently when the satellite is over the leading hemisphere of the
Earth; i.e. when local time is 0 to 12. Actually an insignificant
increase is detected when the satellite is in the lee of the Earth's
movement through the interplanetary medium.
\label{fig11}}
\end{figure}

\begin{figure} \figurenum{12} \caption{The location of the Einstein satellite 
over the Earth at the time of flash detections is indicated by the open
circles. The grey scale map represents the relative exposure time at
that point over the Earth.  The under-exposed area around
$-50{\arcdeg}$ is the location of the South Atlantic Anomaly. The IPC
was turned off when the satellite passed through that region.
\label{fig12}}
\end{figure}

\begin{figure} \figurenum{13} \caption{A grey scale representation of the 
total amount of time the IPC was sensitive to flashes meeting our search
criteria as a function of flash position in detector coordinates.
Note that there is a minimum in the exposure time at the center of the
detector where we are unable to search for faint flashes because of
the typical placement of strong X-ray sources there.  The highest
regions of sensitivity corresponds to the area just outside the
detector window support ribs.
\label{fig13}}
\end{figure}

\begin{figure} \figurenum{14} \caption{The location of flash events 
over the face of the IPC detector. This plot corresponds to the
exposure map in Figure 13. The lines delineate the location of the
detector window support ribs. The outer box encloses the region of the
IPC for which a reliable aspect solution was computed. The flashes are
disproportionately concentrated towards the edge of the detector and
away from the ribs, qualitatively consistent with the exposure map.
\label{fig14}}
\end{figure}

\vfill

\end{document}